\documentclass[aps,pre,twocolumn,showpacs]{revtex4-1}

\usepackage{amsmath,amssymb}
\usepackage{bm}
\usepackage{graphicx}

\usepackage{color}

\newcommand{\tr}{{\rm tr}\,}

\usepackage{ulem}

\begin{document}

\title{
Deformable microswimmer in a swirl: Capturing and scattering dynamics
}

\author{Mitsusuke Tarama$^{1,2,3}$}
\author{Andreas M. Menzel$^2$}
\author{Hartmut L\"owen$^2$}

\affiliation{
$^1$Department of Physics, Kyoto University, Kyoto 606-8502, Japan\\
$^2$Institut f\"ur Theoretische Physik II: Weiche Materie, Heinrich-Heine-Universit\"at D\"usseldorf, D-40225 D\"usseldorf, Germany\\
$^3$Institute for Solid State Physics, University of Tokyo, Kashiwa, Chiba 277-8581, Japan
}

\date{\today}

\begin{abstract}
Inspired by the classical Kepler and Rutherford problem, we investigate an analogous setup in the context of active microswimmers: 
the behavior of a deformable microswimmer in a swirl flow. 
First we identify new steady bound states in the swirl flow and analyze their stability. 
Second we study the dynamics of a self-propelled swimmer heading towards the vortex center, and we observe the subsequent capturing and scattering dynamics. 
We distinguish between two major types of swimmers, those that tend to elongate perpendicularly to the propulsion direction and those that pursue a parallel elongation. 
While the first ones can get caught by the swirl, the second ones were always observed to be scattered, which proposes a promising escape strategy. This offers a route to design artificial microswimmers that show the desired behavior in complicated flow fields. 
It should be straightforward to verify our results in a corresponding quasi-two-dimensional experiment using self-propelled droplets on water surfaces. 
\end{abstract}



\maketitle

\section{Introduction} \label{sec:Introduction}

For many applications it is of key relevance to tune and
control  the motion of artificial and biological microswimmers
\cite{Cates_review, Romanchuk_review,Marchetti_review}
by external influences like confinement \cite{Berke_Lauga_PRL_2008,
Wensink_Loewen_PRE_2008,van_Teeffelen_Loewen_PRE_2008,Zimmermann_Teeffelen_Soft_Matter_2009,Takagi_Palacci_archive,Gompper2013},
solvent flow
\cite{Stocker_PRL,Peyla_2013,Zoettl_Stark_PRL_2012,tenHagen_PRE_2011},
or a magnetic field \cite{Baraban}. This can be exploited to construct
motors and machines on the microscale
\cite{Aranson_PNAS,di_Leonardo_PRL,di_Leonardo_NJP}
and artificial muscles \cite{Juelicher},
to mention just a few examples. In particular, the motion of
self-propelled particles in externally prescribed flow fields gives rise to
significant changes in their swimming paths as shown
by recent studies in planar Couette \cite{tenHagen_PRE_2011} and Poiseuille
flow geometry \cite{Stocker_PRL,Peyla_2013,Zoettl_Stark_PRL_2012}.

Astonishingly the motion of a microswimmer in a swirl has never been considered so far,
although swirl flows occur quite naturally in many situations, including
turbulence. Here we address this problem and augment it
by possible deformations of the particle that couple to the solvent flow.
Using a theoretical description from our earlier work \cite{Tarama2013JCP}, we derive
equations of motion for a swimmer in a swirl. The setup of a swirl
is similar to that of a scattering geometry and possesses therefore an analogy to the
classical Kepler and Rutherford problem. In particular, one can discriminate between the two basic
dynamic events of capturing and scattering: In the former, the swimmer is attracted by the swirl
and cannot escape from it afterwards, while in the latter it escapes from the eddy
by its own self-propulsion. For human swimmers at high Reynolds numbers,
it is important to propose a strategy to escape
a swirl. We address this problem at low Reynolds numbers in an analogous way.
In fact, the two events of capturing, which possibly leads to death, and scattering, corresponding to a successful escape and survival, depend on the impact parameter
and the relative orientation of the swimmer with respect to the flow direction.
In order to discriminate between these two results, we perform a theoretical stability analysis as well as a numerical solution of the corresponding equations of motion.

Our predictions can be tested for deformable swimmers in prescribed vortex flows. Experimental realizations are given by deformable droplets on fluid interfaces propelling due to chemical reactions \cite{Nagai2005,Takabatake2011}. 
The vortex flow profile is of high practical relevance and experimentally easily accessible. In practice, a magnetic stir bar at the bottom of an artificial water tank is enough to steadily maintain it.  Furthermore, it does not require periodic boundary conditions, which are in principle necessary for the typically studied examples of planar Couette or linear shear flow. Nevertheless, we are not aware of any previous investigation of this setup in the presence of active particles. 

This paper is organized as follows: 
in the next section, the time-evolution equations of a deformable microswimmer in a swirl flow are described. 
We discuss the steady-state solutions and their stability in Sec.~\ref{sec:steady solutions}. 
Scattering and capturing dynamics, which actually constitute the central topic of this paper, are considered in Sec.~\ref{sec:capturing and scattering dynamics}, 
together with a brief account of the effect of thermal noise. 
Section \ref{sec:discussion} is devoted to a summary and conclusion. 
Finally, details of the analytical calculation carried out in Sec.~\ref{sec:steady solutions} are explained in the appendix.

\section{Model} \label{sec:Model}

In the following, we introduce the model equations for an active deformable microswimmer in a swirl flow. 
We consider a two-dimensional environment and denote the fluid flow field as $\mathbf{u}(x,y)$. 
Our simple vortex flow (swirl) is given by 
\begin{equation}
\mathbf{u}(x,y) = \left( \mu \frac{-y}{x^2+y^2}, \mu \frac{x}{x^2+y^2} \right),
 \label{eq:u}
\end{equation}
where $\mu$ sets the strength of the vortex. 
It describes a rotational flow around a swirl center. 
Naturally, the swirl center defines the origin of our coordinate frame. A flow potential exists for this type of fluid flow such that $\mathbf{u}(x,y)=-\nabla U(x,y)$, where $U(x,y)=\mu\arctan(x/y)$. 
Consequently, $\nabla\times\mathbf{u}=\mathbf{0}$, which implies that the fluid flow does not contain a local rotational contribution. 
This swirl flow can easily be realized in an experimental setup. Surface swimmers on a fluid provide a reasonable realization of our two-dimensional considerations. 

For a simple swimmer, two basic kinds of activity can generally be distinguished: a spontaneous translational motion (self-propulsion) and a spontaneous rotation (spinning motion). The impact of spontaneous active rotations has been investigated in detail \cite{Tarama2012,Tarama2013PTEP,Tarama2013PRE,Wittkowski2012,tenHagen_PRE_2011,Zoettl_Stark_PRL_2012,Fily-Baskaran-Marchetti2012,Glotzer2013,Uchida2010,Uchida2011,Kummel2013,Kapral2010,Takagi2013}. 
Here, for simplicity, we take into consideration only a spontaneous translational motion, i.e., self-propulsion. 
The time-evolution equations for a deformable active particle in a fluid flow field $\mathbf{u}$ then can be derived from symmetry arguments as \cite{Tarama2013JCP}
\begin{equation}
\frac{d x_i}{dt}= u_i +v_i,
\label{eq:x}
\end{equation}
\begin{equation}
\frac{d v_i}{dt} 
= \gamma v_i -\left( v_k v_k \right) v_i -a_1 S_{ik}v_k, 
\label{eq:v}
\end{equation}
\begin{equation}
\frac{d S_{ij}}{dt} 
= -\kappa S_{ij}+b_1\!\! \left[ v_i v_j -\frac{\delta_{ij}}{2} \left( v_k v_k \right) \right] 
 + \nu_1\!\! \left[ A_{ij}-\frac{\delta_{ij}}{2} A_{kk} \right]\!\!, 
 \label{eq:S}
\end{equation}
where $\delta_{ij}$ is the Kronecker delta and 
summation over repeated indices is implied. 

In the above equations, $\mathbf{x}=(x_1,x_2)$ represents the position of the center of mass of the swimmer. 
It is parameterized as 
\begin{equation}
\mathbf{x} = \left( r \cos \eta, r \sin \eta \right)
 \label{eq:x2d}
\end{equation}
in polar coordinates. 
When viewed from the laboratory frame as in Eq.~(\ref{eq:x}), the microswimmer in total moves with 
the velocity $\mathbf{u} +\mathbf{v}$. In this expression, $\mathbf{u}$ is the imposed flow velocity of the fluid given by Eq.~(\ref{eq:u}), while $\mathbf{v}$ is the relative swimming velocity of the swimmer with respect to its fluid environment. We parameterize $\mathbf{v}$ as
\begin{equation}
\mathbf{v} = \left( v \cos \phi, v \sin \phi \right).
 \label{eq:velocity}
\end{equation}

The time evolution of the relative velocity $\mathbf{v}$ is given by Eq.~(\ref{eq:v}). 
On the right-hand side, the coefficient $\gamma$ would generally be negative for passive particles and describe the friction with the fluid environment. 
For active swimmers, however, it becomes positive, $\gamma>0$. Together with the stabilizing cubic velocity term in Eq.~(\ref{eq:v}), a nonvanishing relative velocity $\mathbf{v}\neq\mathbf{0}$ of active swimming becomes possible. The direction of $\mathbf{v}$ is not fixed and results from spontaneous symmetry breaking. This is a key ingredient to active self-propelled motion in the absence of external fields. 
An analogous approach had already been established by the early continuum descriptions of flocks of active particles \cite{TonerTuPRL, TonerTuPRE, TonerTuRamaswamy}. 

Next, the last term in Eq.~(\ref{eq:v}) with the coefficient $a_1$ includes the leading-order coupling of the velocity to deformations $\mathbf{S}$ of the swimmer. The influence of this term on the swimmer behavior will be addressed below together with its counterpart in Eq.~(\ref{eq:S}), the term with the coefficient $b_1$. 

\begin{figure}
  \begin{center}
         \includegraphics[width=0.4\textwidth]{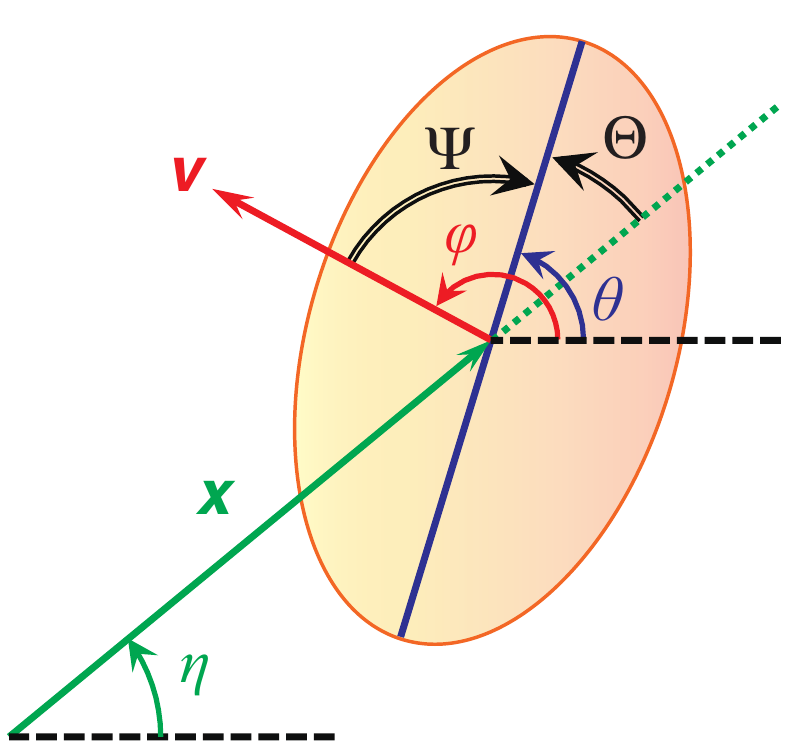}
      \caption{
      Variables characterizing the orientational degrees of freedom of an elliptically deformable active microswimmer when viewed from the laboratory frame (schematic). 
      } \label{fig:angles}
  \end{center}
\end{figure}

Here we take into account only elliptic deformations of the swimmer in a lowest-order approach. As a consequence, the tensor $\mathbf{S}$ is of second rank, symmetric, and traceless. 
It can be directly related to the second Fourier mode of the shape changes of a deformable particle \cite{OOS2009}. Furthermore, it 
is of the same form as the order parameter used to characterize the state of nematic liquid crystals \cite{degennes}. 
Its components are parameterized as 
\begin{gather}
S_{11}=-S_{22}=s \cos 2 \theta, \notag\\
S_{12}=S_{21}=s \sin 2\theta.
 \label{eq:S2d}
\end{gather}
Here $s$ characterizes the degree of elliptic deformation, while $\theta$ measures the orientation of the symmetry axis of elliptic deformation when viewed from the laboratory frame. See Fig.~\ref{fig:angles} for the assignment of the angles $\eta$, $\phi$, and $\theta$. 

We consider the time evolution of the deformations $\mathbf{S}$ as described by Eq.~(\ref{eq:S}). 
First, we set $\kappa>0$ in the leading term on the right-hand side. This implies that 
we do not include active deformations of the swimmer. Its shape rather tends to relax back to the undeformed state when the other contributions are ignored. Thus in the limit of $\kappa\rightarrow\infty$ we obtain $\mathbf{S}=\mathbf{0}$ and our equations apply to the case of a rigid spherical swimmer. 

Next, by the contribution with the coefficient $b_1$, we include the leading-order coupling to the relative velocity. To maintain the requirements on the structure of $\mathbf{S}$, i.e., that this tensor be symmetric and traceless, the coupling term is constructed in a symmetric and traceless form as well. We come back to the influence of this term below. 

Finally, the tensor $\mathbf{A}$ in Eq.~(\ref{eq:S}) describes the strain rate (dynamic elongational contribution) 
due to the fluid flow. Its components are given by  
\begin{equation}
A_{ij} = \frac{1}{2} \left( \partial_i u_j +\partial_j u_i \right). 
\label{eq:A}
\end{equation}
$\mathbf{A}$ itself is already symmetric, but its contribution is rendered traceless in Eq.~(\ref{eq:S}) to keep the traceless nature of $\mathbf{S}$. 
A corresponding term is already present in the characterization of passive deformable objects \cite{Maffettone1998}. 
The coefficient $\nu_1$ determines how strongly the elongational part of the fluid flow stretches the deformable swimmer. 

Our set of equations of motion Eqs.~(\ref{eq:x})--(\ref{eq:S}) 
was derived from symmetry arguments and not on the basis of a specific microscopic model. This serves to keep the description as general as possible. Only the leading-order coupling terms between velocity and deformation as well as to the external flow field are included. 
The description reduces to known models in several limiting cases such as for a passive deformable particle in shear flow \cite{Maffettone1998}, an active rigid particle in shear flow \cite{Wittkowski2012}, and an active deformable particle in a quiescent environment \cite{OhtaOhkuma2009,Tarama2012,Tarama2013PTEP}. 
Not including any external flow field, Eqs.~(\ref{eq:x})--(\ref{eq:S}) were derived theoretically for an isolated deformable domain in a reaction-diffusion system \cite{OOS2009,Shitara2011}. 

We now come back to the leading coupling terms between the relative velocity $\mathbf{v}$ and the deformation $\mathbf{S}$ described by the coefficients $a_1$ and $b_1$. 
Their impact has been studied in detail in the absence of external flow fields 
\cite{OhtaOhkuma2009,Tarama2011,Hiraiwa2010,Hiraiwa2011,Itino2011,Itino2012,Menzel2012,epjst}.  
They in general imply that propulsion can lead to shape changes, and in turn deformations can influence the propulsion direction and the swimming speed. 
One important consequence of these terms is that increased deformations lead to curved trajectories. Without the fluid flow, a bifurcation from straight to circular motion is found from Eqs.~(\ref{eq:v}) and (\ref{eq:S}). It occurs at a critical value $\gamma=\gamma_c$, where \cite{OhtaOhkuma2009}
\begin{equation}
\gamma_c  = \frac{\kappa^2}{a_1 b_1} +\frac{\kappa}{2}.
 \label{eq:gamma_c}
\end{equation}
The signature of this bifurcation will also affect the behavior of our swimmer in the swirl flow below. 

Besides influencing the dynamics, the coefficients $a_1$ and $b_1$ encode aspects of the swimmer geometry. 
On the one hand, if $a_1<0$ and $b_1<0$, the symmetry axis of the elliptical deformation tends to be perpendicular to the propulsion velocity (``perpendicular case'') \cite{OhtaOhkuma2009}. On the other hand, for $a_1>0$ and $b_1>0$, the two directions tend to be parallel to each other (``parallel case'') \cite{OhtaOhkuma2009}. 
Experimentally a corresponding example system is given by a deformable droplet self-propelling on a fluid interface due to a chemical reaction
\cite{Nagai2005}. 
A recent theoretical investigation \cite{Yoshinaga2013} suggests that such a system is a pusher in the perpendicular case and a puller in the parallel case. 
$a_1$ and $b_1$ are thus determined by the relation between propulsion and deformation that in turn results from the microscopic propulsion mechanism. 
For an explicit calculation of these coefficients in a microscopic model of self-propelled droplets see Ref.~\cite{Yoshinaga2013}.

\section{Steady-state solutions} \label{sec:steady solutions}

The swirl flow in Eq.~(\ref{eq:u}) features a rotational symmetry. We therefore expect that circular steady-state solutions of closed circular loops exist, at least for passive particles. 
For rigid spherical passive particles, implying $\gamma<0$ and $\kappa\rightarrow\infty$, Eqs.~(\ref{eq:x})--(\ref{eq:S}) reduce to $d\mathbf{x}/dt=\mathbf{u}$, $\mathbf{v}=\mathbf{0}$, and $\mathbf{S}=\mathbf{0}$. Thus these objects are simply advected by the fluid flow on circles around the vortex center. The stability of the circular motion of any radius is marginal, with the radius only determined by the initial conditions. 
In the following, we 
investigate how deformability and self-propulsion 
change this result. 
For this purpose, we analyze the stability of the steady-state solutions of Eqs.~(\ref{eq:x})--(\ref{eq:S}). 

\subsection{Passive circular motion of deformable particles} \label{subsec:passive circular motion}

We first consider a deformable passive particle, i.e.\ one that is not self-propelled. 
In this case $\gamma<0$. 
Consequently it is simply convected by the fluid flow, and the relative velocity with respect to the surrounding fluid vanishes, $\mathbf{v}=\mathbf{0}$. Nevertheless, the particle can be deformed by the elongational component of the flow field. 

Under these assumptions, we investigate the steady-state solutions of Eqs.~(\ref{eq:x})--(\ref{eq:S}). 
As explained in more detail in the appendix, we find circular trajectories of fixed radius $r=r_0$. Generally, the particle is deformed. Its deformation axis is tilted by a constant angle ($\Theta$ in the appendix) with respect to the radial direction of the circle. 
An illustrative picture of these trajectories is given by a permanently deformed particle, anchored under a constant angle to the rotating pointer of a clock. 
We refer to this situation as {\it passive circular motion}.

If the radius $r_0$ of the circle is below a certain value, $r_0 < r_{0, {\rm min}}$, all passive circular motions $\gamma<0$ are marginally stable. Above this value, $r_0 \ge r_{0, {\rm min}}$,  trajectories are marginally stable only for stronger friction $\gamma<\gamma_{-}<0$. This is illustrated in Fig.~\ref{fig:gamma-vsr0} with the detailed analysis given in the appendix. The stability is marginal with respect to the radial direction. It is asymptotic with respect to all other degrees of freedom. 

We tested and supplemented these analytical results by numerically solving the equations of motion Eqs.~(\ref{eq:x})--(\ref{eq:S}). We always employed a fourth order Runge-Kutta method of time increment $\delta t = 10^{-3}$ and checked our results using still finer time steps. Here we initialized the system by circular particles placed at different distances from the vortex center. 

Figure \ref{fig:gamma-vsr0} shows corresponding numerical results for the system parameters chosen as $\kappa=0.5$, $a_1=b_1=-1$, $\nu_1=1$, and $\mu=1$, i.e., $\kappa < \left| a_1 \nu_1 \right|$. This implies soft deformable passive particles.
\begin{figure}
  \begin{center}
         \includegraphics[width=0.4\textwidth]{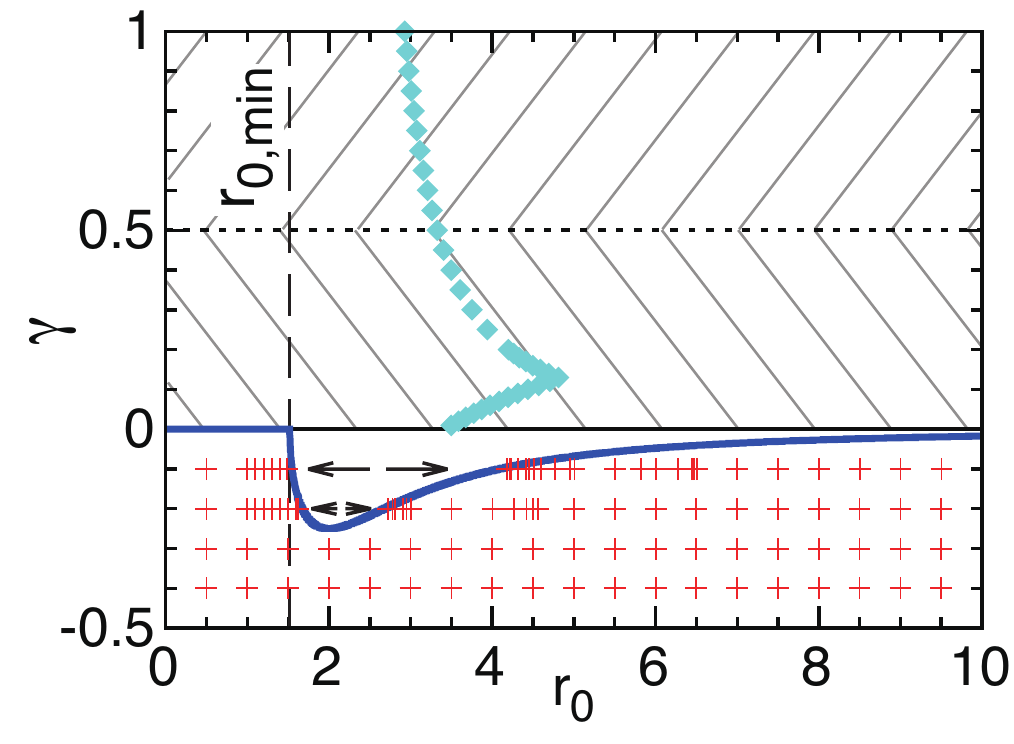}
      \caption{
      Regions of stable circular motions for soft deformable particles 
      in the perpendicular configuration. 
      $r_0$ gives the radius of the circular trajectory. 
      $\gamma>0$ measures the self-propulsion speed of active swimmers (diamonds), 
      whereas $\gamma<0$ characterizes the friction for passive particles (plusses). 
      We indicate by the vertical dashed line the value of the 
      radius $r_{0, {\rm min}}$, see main text and Eq.~(\ref{eq:r0min}). 
      The solid (blue) line gives the stability range of circular trajectories 
      for deformable passive particles. This line can be calculated analytically, see Eqs.~(\ref{eq:gamma-}) and (\ref{eq:stability_passive_circular}). 
      Passive systems initialized in the ``dip'' region are horizontally expelled from 
      that area as indicated by the arrows. 
      Active swimmers perform different types of motion, 
      depending on the initial conditions: 
      an active circular motion (diamonds)
      or an escape (not indicated) for $0<\gamma<0.5$ (lower hatched area); 
      an active circular motion (diamonds) or a lunar-type motion (not indicated) 
      for $\gamma>0.5$ (upper hatched area). 
      The system parameters are set to $\kappa=0.5$, $a_1=b_1=-1$, $\nu_1=1$, and $\mu=1$.
      } \label{fig:gamma-vsr0}
  \end{center}
\end{figure}
The solid (blue) line 
marks the analytical stability limit, cf.\ Eqs.~(\ref{eq:gamma-}) and (\ref{eq:stability_passive_circular}). 
Circular trajectories of radius $r_0$ with a value of $\gamma$ below this line are stable. 
On the contrary, if we initialize trajectories with $\gamma<0$ above the solid stability line (i.e., within the ``dip'' region in Fig.~\ref{fig:gamma-vsr0}), the system is expelled from that area, as indicated by the horizontal arrows. Most of the dense points in Fig.~\ref{fig:gamma-vsr0} on both sides of the dip follow from such expelled systems. As we can see, the expelled systems typically overshoot the stability line until they finally get stabilized. We checked numerically that the results are qualitatively the same 
when $a_1=b_1=+1$, as predicted from our theoretical analysis.

\subsection{Active circular motion of deformable swimmers} \label{subsec:Active circular motion}

We now turn to steady-state active motions of $\mathbf{v} \neq \mathbf{0}$ for $\gamma > 0$. 
As noted above, in the absence of the fluid flow a straight active motion is obtained for $0<\gamma<\gamma_c$ and a circular motion for $\gamma>\gamma_c$, with $\gamma_c$ given in Eq.~(\ref{eq:gamma_c}) \cite{OhtaOhkuma2009}.
In the presence of the fluid flow, we solved the dynamic equations (\ref{eq:x})--(\ref{eq:S}) numerically. 
As an initial condition, the variables were set according to the analytical solution of 
active straight motion 
without fluid flow \cite{OhtaOhkuma2009}, and the swimmer was placed at various distances $r_{\rm init}$ from the vortex center. 
We chose the same system parameters as mentioned in the caption of Fig.~\ref{fig:gamma-vsr0}, 
except that  
we considered both cases of $a_1 = b_1 = \mp1$. 
As mentioned above, in the absence of the fluid flow, the deformation axis tends to be perpendicular to the propulsion direction for $a_1 = b_1 = -1$,  whereas it takes a parallel configuration for $a_1 = b_1 = +1$ \cite{OhtaOhkuma2009}. 

On the one hand, for the perpendicular case ($a_1 = b_1 = -1$) 
we find a steady-state {\it active circular motion} when $0<\gamma<\gamma_c$, where here $\gamma_c=0.5$. It is the analog of the passive circular motion of $\gamma<0$. However, in contrast to the passive case, we now obtain only one stable diameter $r_0$ for each value of $0<\gamma<\gamma_c$. This is indicated by the diamond symbols in Fig.~\ref{fig:gamma-vsr0}. 
Depending on the initial conditions, the swimmer 
either asymptotically approaches this orbit, or it manages to escape from the swirl to infinite distance.

\begin{figure*}
  \begin{center}
         \includegraphics[width=0.8\textwidth]{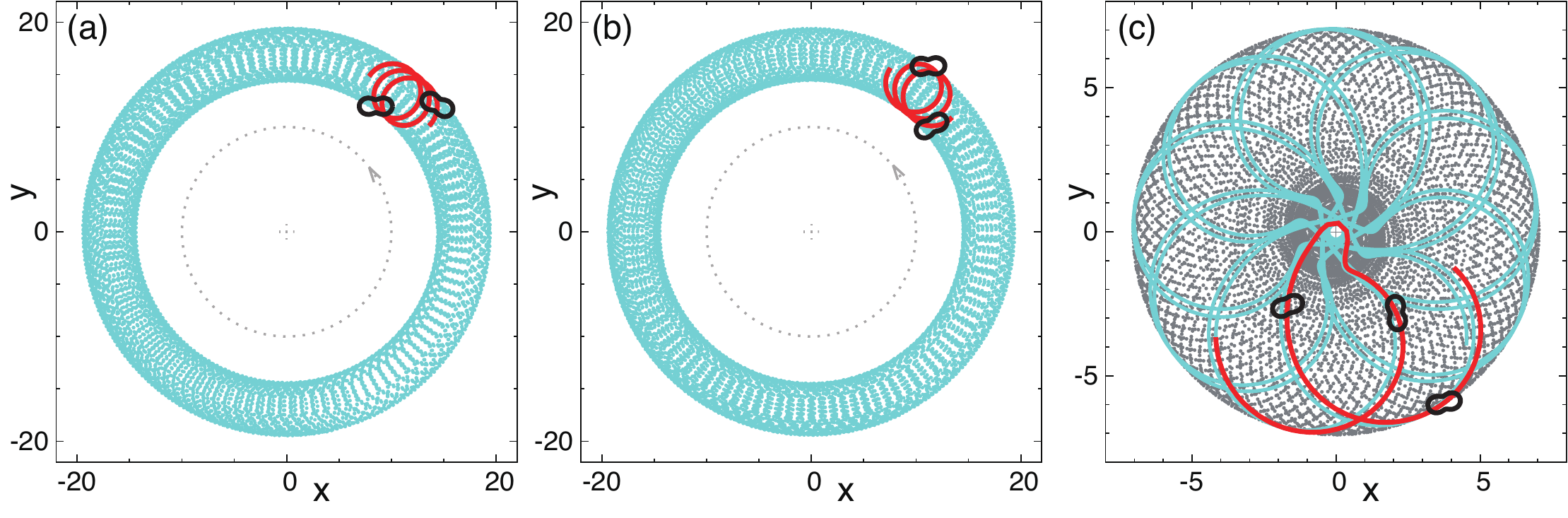}
      \caption{
      Example trajectories of (a) and (b) {\it lunar-type motions} and (c) a {\it multi-circular motion}, 
      each for $\gamma=1$ ($>\gamma_c$). 
      Further parameter values are (a) $a_1 = b_1 = -1$ (perpendicular case) and $r_{\rm init}=10$, as well as (b) and (c) $a_1=b_1=+1$ (parallel case) together with (b) $r_{\rm init}=10$ and (c) $r_{\rm init}=0.5$. 
      The remaining parameters are set to $\kappa=0.5$, $\nu_1=1$, and $\mu=1$.
      Different colors mark trajectory pieces of different intervals of swimming time. Black superimposed silhouettes indicate particle orientations and degrees of deformation.
      } \label{fig:quasi-periodic}
  \end{center}
\end{figure*}

For $\gamma > \gamma _c$, the situation becomes markedly different. Starting sufficiently close to the radius $r_0$, we still observe the steady-state active circular motion as indicated in Fig.~\ref{fig:gamma-vsr0}. However, another type of motion occurs depending on the initial conditions. 
We call it a {\it lunar-type motion} and depict a typical trajectory in Fig.~\ref{fig:quasi-periodic}(a) for $\gamma = 1$. 
A short-time piece of the trajectory is emphasized by the thick solid (red) line. The black superimposed silhouettes show the particle orientations and degrees of deformation in an exaggerated way for illustration. We can understand this trajectory as the circular motion that already occurs in the absence of the swirl for $\gamma>\gamma_c$ \cite[Eq.~(\ref{eq:gamma_c})]{OhtaOhkuma2009} superimposed onto the circular convection due to the vortex flow. 
In this case, both rotational directions---the one of the smaller revolution, corresponding to the circular trajectory of the moon around the Earth in our heliocentric picture, and the one of the larger revolution, corresponding to the trajectory of the Earth around the sun---have the same sense of rotation as the fluid flow. 
The radius of the larger revolution depends on the initial condition. 

On the other hand, in the parallel case ($a_1 = b_1 = +1$) 
we did not observe a steady-state active circular motion.
Instead, all particles escape far from the vortex center for $0 < \gamma < \gamma_c = 0.5$. 
Therefore designing an active swimmer in the parallel configuration at low propulsion speed offers a promising strategy to allow escapes. 
In contrast, for $\gamma > \gamma_c$, a particle again undergoes a lunar-type motion. We display 
a typical trajectory in Fig.~\ref{fig:quasi-periodic}(b). 
Here, however, the smaller revolution and the fluid flow have opposite sense of rotation, whereas the larger revolution and the fluid flow share the same sense of rotation. 
This differs from the perpendicular case 
considered above.  
Again, the radius of the larger revolution depends on the initial condition. 

With increasing $\gamma \geq 0.7$, the situation gets still more complex 
in the parallel case. 
Depending on the initial condition, a {\it multi-circular motion} 
can emerge as illustrated in Fig.~\ref{fig:quasi-periodic}(c). 
The lighter gray, thicker solid (turquoise), and thick solid (red) lines show trajectory pieces of decreasing swimming-time intervals. 
To obtain the multi-circular motion, the swimmer was initially placed relatively close to the vortex center. 

In summary, these observations suggest the following escape strategy for a deformable swimmer that cannot actively determine its swimming direction and was dragged into the swirl. 
If possible, a parallel configuration should be adopted. 
(Within the framework of Ref.~\cite{Yoshinaga2013} this corresponds to a puller-like propulsion mechanism.) 
Then the most effective way is not to try too hard to escape; 
in other words, the effort of self-propulsion should be kept low ($\gamma<\gamma_c$). In this combined situation we always observed that the swimmer manages to escape.

\section{Capturing and scattering dynamics} \label{sec:capturing and scattering dynamics}

\begin{figure}
  \begin{center}
         \includegraphics[width=0.4\textwidth]{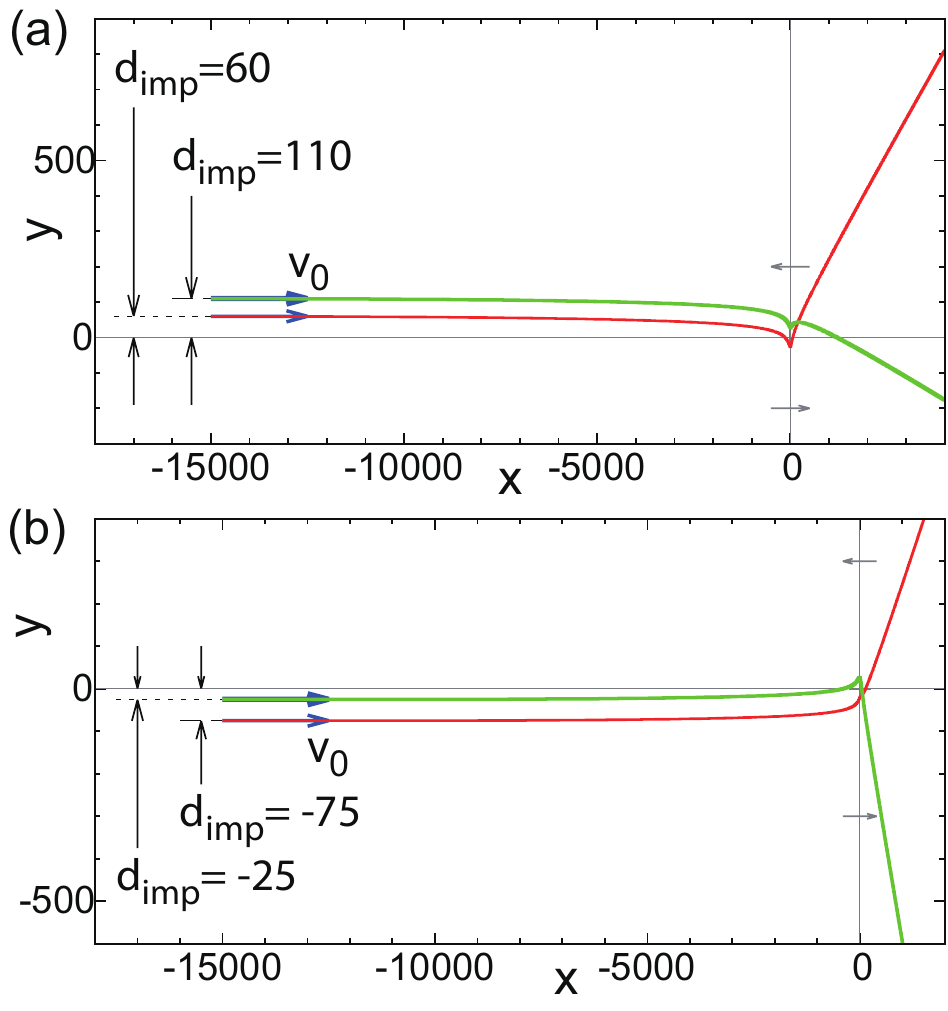}
      \caption{
      Definition of the impact parameter $d_{\rm imp}$ depicted using example trajectories of incident swimmers (a) in a perpendicular configuration ($a_1=b_1=-1$) and (b) in a parallel configuration ($a_1=b_1=1$) for $\gamma=0.3$. The impact parameter $d_{\rm imp}$ measures the distance by which the swimmer would miss the swirl center, if its trajectory were not affected by the swirl flow. 
      Gray arrows on the axis $x=0$ indicate the direction of the swirl flow, pointing to the left for $y>0$ and to the right for $y<0$. The sign of $d_{\rm imp}$ is defined as positive when the swimmer initially heads towards the side $y>0$ of oppositely directed fluid flow; it is chosen negative when the swimmer initially heads towards the side $y<0$ of identically oriented fluid flow. 
      For illustration, the scales of the $x$ and $y$ axes are chosen differently and distances in the $y$ direction are enlarged by a factor of ten. 
      } \label{fig:incident_trajectory}
  \end{center}
\end{figure}

We now study the ``collision'' of an active deformable swimmer with the swirl. 
This is performed in analogy to a classical scattering experiment. Therefore the swimmer is initially not placed close to the vortex center, but at a comparatively large distance $r_{\rm init}$ away. 
If the swimmer were not affected by the flow field of the swirl, it would propel along the direction of its initial velocity orientation. The distance $d_{\rm imp}$ by which it would then miss the vortex center is called the impact parameter. 
A swimmer of $d_{\rm imp}=0$ would hit the center of the vortex, if it were not affected by the swirl flow. 
The definition of $d_{\rm imp}$ is illustrated in Fig.~\ref{fig:incident_trajectory} using example trajectories that we will discuss in more detail below. 

If the particle manages to escape from the vortex, we can measure the scattering angle of the event. For this purpose, we determine the angle $\eta_{\rm scat}$ between the initial velocity orientation and the final velocity orientation when the particle has reached a certain distance $r_{\rm esc}$ from the vortex center. 

To keep the setup simple and meaningful in the sense of a scattering experiment, we set the propulsion strength to values $0<\gamma<\gamma_c$. For these values, an active straight motion occurs in the absence of the flow field \cite{OhtaOhkuma2009}. We provide this solution as an initial condition at a very large initial distance $r_{\rm init}=1.5 \times10^4$. 
After numerically integrating Eqs.~(\ref{eq:x})--(\ref{eq:S}) forward in time, we measure the scattering angle at the distance $r_{\rm esc}=10^{4}$ if a scattering event occurs. We varied the values of the propulsion strength $\gamma$ and the impact parameter $d_{\rm imp}$, while the other parameters were chosen as before. 
Due to the swirl geometry, the events of passing the vortex center on one side differs from passing it on the other side. We therefore define $d_{\rm imp}>0$ when the particle velocity is initially oriented towards the side of oppositely directed fluid flow. On the contrary, we set $d_{\rm imp}<0$ when the swimmer initially propels towards the side of identically directed fluid flow. 
See Fig.~\ref{fig:incident_trajectory} for an illustration of the definition of the sign of $d_{\rm imp}$. 

We systematically initialized and analyzed scattering events as a function of varying impact parameter $d_{\rm imp}$. Our results are presented in the following, with example trajectories displayed in Figs.~\ref{fig:perpendicular} and \ref{fig:parallel}. As can be inferred from the different axes scales of Figs.~\ref{fig:perpendicular} and \ref{fig:parallel} in comparison to those of Fig.~\ref{fig:incident_trajectory}, the trajectory parts shown in Figs.~\ref{fig:perpendicular} and \ref{fig:parallel} represent a zoom onto the close vicinity around the swirl center. In this area, the trajectories have already been significantly bent and influenced by the swirl flow, which becomes obvious from the comparison to Fig.~\ref{fig:incident_trajectory}. 
Again, we distinguish between a perpendicular configuration ($a_1=b_1=-1$) 
and a parallel configuration ($a_1=b_1=+1$).

\subsection{Perpendicular case} \label{sec:perpendicular}

For the perpendicular configuration ($a_1=b_1=-1$), the particle tends to orient its deformation axis perpendicularly to the propulsion velocity. 
As we saw in the previous section, this is not the best strategy to escape from the vortex. 
Indeed, as we will see shortly, in a finite interval of impact parameters, the particle gets captured by the swirl instead of being scattered. We discuss the situation as a function of increasing impact parameter. 

Generally, the vortex makes the swimmer deviate from its straight trajectory of motion, see Fig.~\ref{fig:incident_trajectory}. 
\begin{figure}
  \begin{center}
         \includegraphics[width=0.425\textwidth]{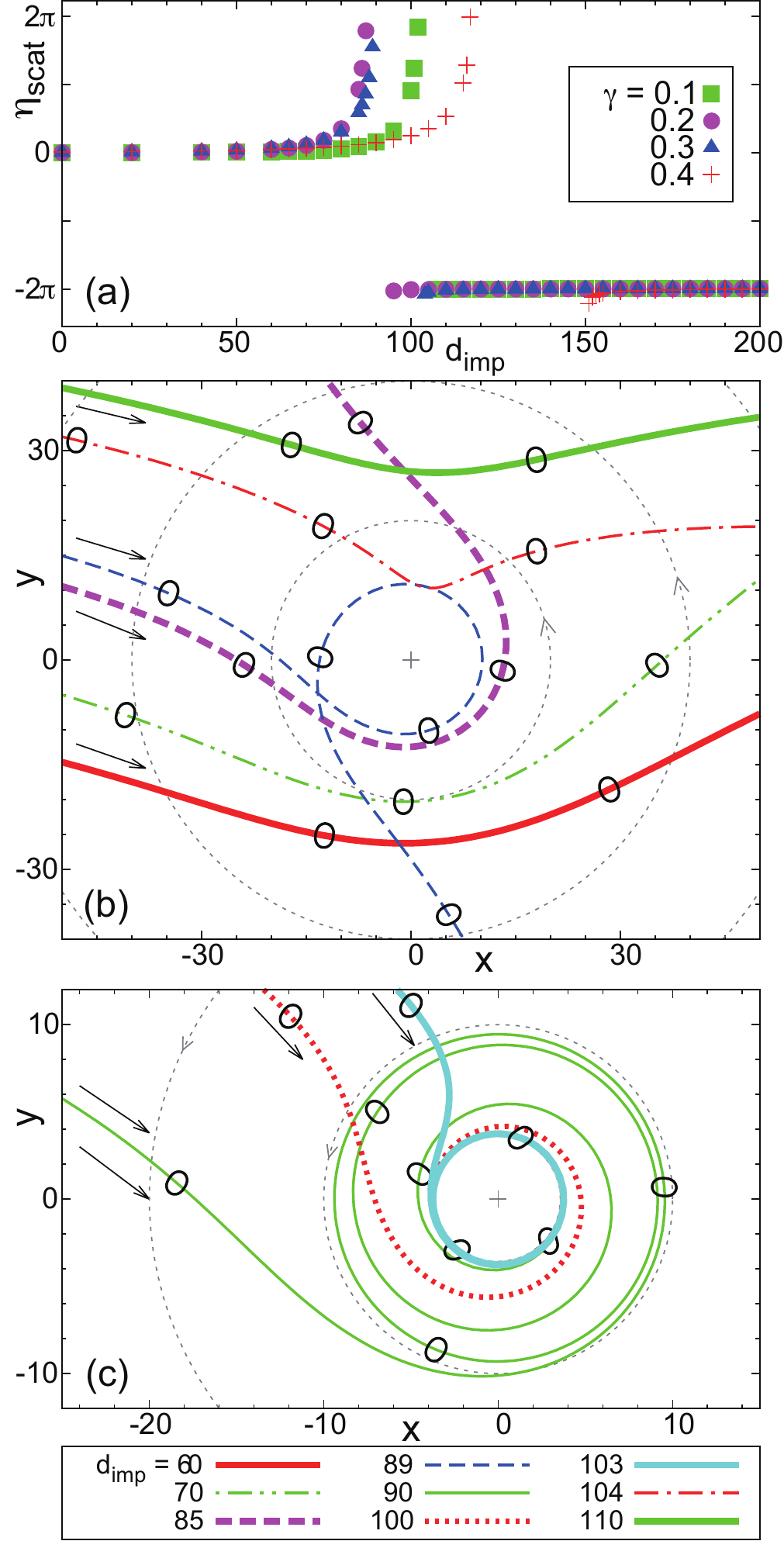}
      \caption{
      Scattering dynamics for $a_1=b_1=-1$ (perpendicular case): (a) scattering angle $\eta_{\rm scat}$ as a function of the impact parameter $d_{\rm imp}$ for various propulsion strengths $\gamma$; 
      (b,c) example trajectories for $\gamma=0.3$ and for different impact parameters $d_{\rm imp}$.
      Black superimposed silhouettes indicate the swimmer orientations and degrees of deformation. 
      Arrows mark the direction of motion. 
      Furthermore, the gray dotted lines illustrate the direction of the fluid flow, the vortex center marked by the plus symbol. (c) There is a finite interval of impact parameters for which the swimmer cannot escape from the vortex any more but gets captured by the swirl. 
      (Other parameter values are $\kappa=0.5$, $\nu_1=1$, and $\mu=1$.)
      } \label{fig:perpendicular}
  \end{center}
\end{figure}
For negative impact parameters $d_{\rm imp}$ (not displayed), the particle trajectory is only weakly deformed and the particle leaves the vortex geometry with basically the same velocity orientation that it had when entering the setup. Thus the scattering angle $\eta_{\rm scat}$ vanishes. This is true even for weakly positive impact parameters. 
We recall at this point that the magnitude of the impact parameter and its sign are defined for a quiescent reference situation, measuring by how much and on which side the swimmer would miss the center in the absence of the swirl flow, respectively. 
In the presence of the vortex flow, however, the swirl can guide the swimmer around the vortex center even on the side opposite to the one that the swimmer is initially heading towards. 
Still, the net scattering angle $\eta_{\rm scat}$ can be relatively small in this case. 
Such a situation is illustrated in Figs.~\ref{fig:incident_trajectory}(a), \ref{fig:perpendicular}(a), and \ref{fig:perpendicular}(b) for $\gamma=0.3$ and $d_{\rm imp} = 60$. The scattering angle appears higher in Fig.~\ref{fig:incident_trajectory}(a) due to the rescaled $y$ dimension; see Fig.~\ref{fig:perpendicular}(a) for the absolute value. 

Further increasing the impact parameter, the scattering process becomes more persistent. 
Now the particle reaches closer to the vortex center and its trajectory gets significantly bent. 
This results in 
a nonzero scattering angle $\eta_{\rm scat}$, see Fig.~\ref{fig:perpendicular}(a) and the trajectories for $d_{\rm imp} = 70$ and $85$ in Fig.~\ref{fig:perpendicular}(b). 

Then, as a function of increasing impact parameter, the swimmer more and more gets caught by the swirl. It can happen that the swimmer circles around the vortex center before it can finally escape from the swirl as depicted by the trajectory for $d_{\rm imp} = 89$ in Fig.~\ref{fig:perpendicular}(b). These events correspond to reorientation processes that are more extreme than simple backscattering, and we indicate them by scattering angles $\eta_{\rm scat}>\pi$ in Fig.~\ref{fig:perpendicular}(a). The scattering angle seems to diverge when the impact parameter is further increased. 

At still higher impact parameters, the swimmer eventually gets captured by the vortex. It cannot escape from the swirl any more. 
Example trajectories are depicted in Fig.~\ref{fig:perpendicular}(c) for $d_{\rm imp}=90$, $100$, and $103$. 
Interestingly, in all these cases the swimmer ends up on the same circular trajectory around the vortex center. This attractant type of motion corresponds to the active circular motion discussed in Sec.~\ref{subsec:Active circular motion}. 

Finally, when the impact parameter is too large, the swimmer does not get close enough any more to the swirl center to be effectively captured. It now gets scattered again, passing the vortex center on the other side, however. To identify these events of passing on the other side of the swirl in Fig.~\ref{fig:perpendicular}(a), we shifted the corresponding scattering angles $\eta_{\rm scat}$ by $-2\pi$. Such events are illustrated in Fig.~\ref{fig:perpendicular}(b) by the trajectories for $d_{\rm imp}=104$ and $110$. 
The complete trajectory of the scattering event for $d_{\rm imp}=110$ is depicted in Fig.~\ref{fig:incident_trajectory}(a).

In effect, we found that the swimmer gets scattered by the swirl and can escape for low-enough impact parameters $d_{\rm imp}$. It gets captured by the swirl at intermediate impact parameters. For large-enough impact parameters, it gets scattered again and can escape. Thus the dynamical behavior of getting scattered is {\it reentrant} as a function of the impact parameter $d_{\rm imp}$. 
\begin{figure}
  \begin{center}
         \includegraphics[width=0.4\textwidth]{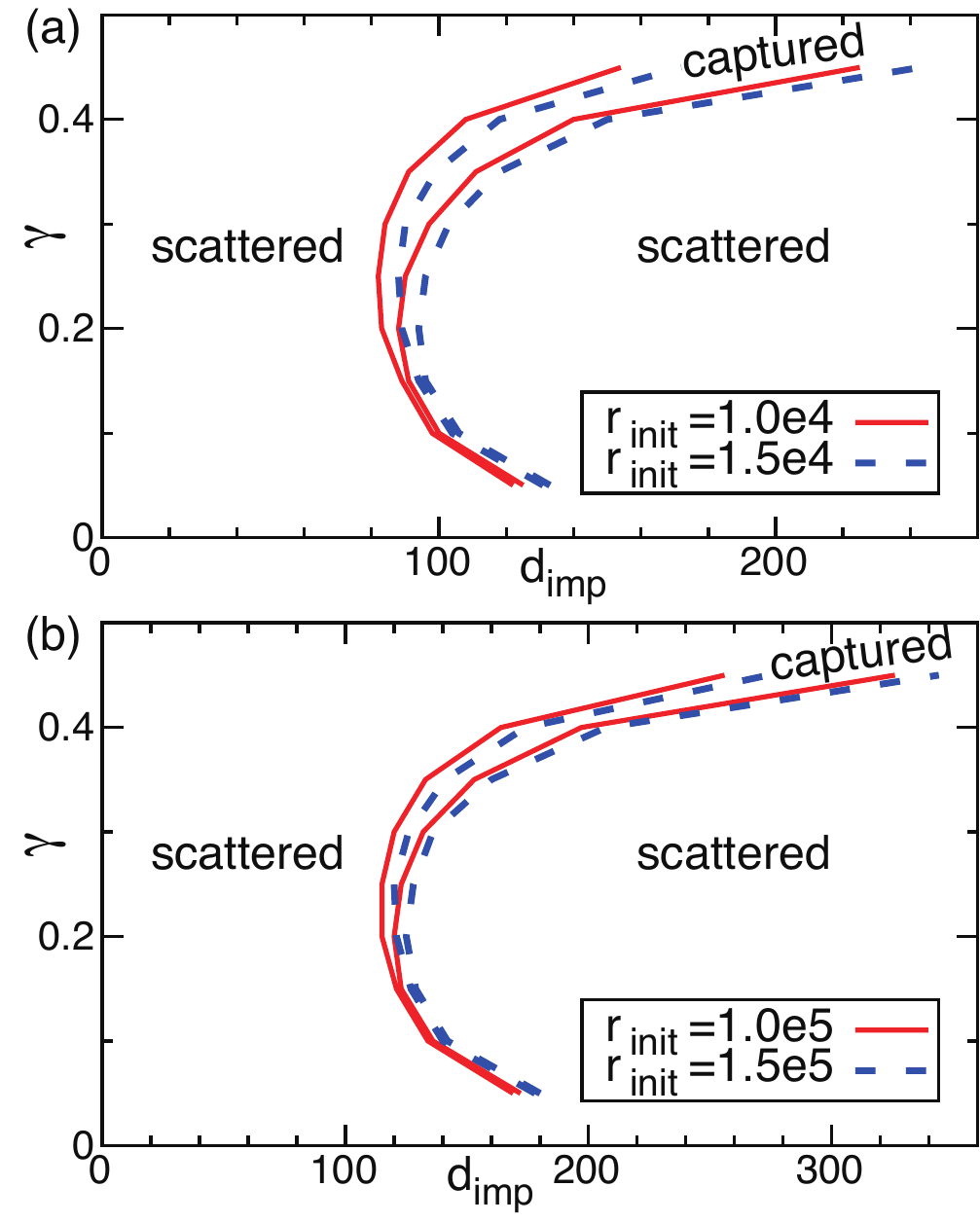}
      \caption{
      Dynamical phase diagram as a function of the impact parameter $d_{\rm imp}$ 
      and the strength of active propulsion $\gamma$
      for the perpendicular case ($a_1=b_1=-1$). 
      Two dynamic events are distinguished: 
      getting scattered by the swirl with a subsequent escape 
      (area outside the lines); 
      and getting captured by the vortex 
      (area between the lines). 
      Our results do not qualitatively depend on the initial distance 
      of the swimmer from the vortex center as can be seen when comparing 
      panels (a) and (b) with each other, or when comparing the solid lines with the 
      dashed lines within each panel. 
      Reentrance of the dynamic events can be observed in both directions 
      of the parameter space. 
      } \label{fig:impact_factor}
  \end{center}
\end{figure}
Guided by this observation, we scanned the swimmer behavior in the parameter plane of the impact parameter $d_{\rm imp}$ and the active propulsion strength $\gamma$. We distinguished between events of scattering and escape on the one hand and events of capturing on the other hand. The resulting dynamic phase diagram is shown in Fig.~\ref{fig:impact_factor}. 
Most interestingly, the phase behavior is not only reentrant as a function of the impact parameter $d_{\rm imp}$ for fixed propulsion strength $\gamma$. Rather, at fixed intermediate impact parameter $d_{\rm imp}$, we also observe reentrance of the capturing event and a twofold reentrance of the scattering behavior with increasing propulsion strength $\gamma$. 
We checked that our results only slightly vary with the initial distance $r_{\rm init}$ from the vortex center due to the spatial decay of the flow field. 
Qualitatively our results do not depend on the initial distance $r_{\rm init}$.

\subsection{Parallel configuration} \label{sec:parallel}

For the parallel configuration ($a_1=b_1=+1$), the swimmer tends to orient its deformation axis along the direction of self-propulsion. This is a good strategy to avoid getting captured by the swirl. Indeed we never observed any event of permanent capturing for such swimmers that started far from the vortex center with propulsion strengths $\gamma<\gamma_c$. 
Again, we discuss the changes in the dynamic behavior with increasing impact parameter. 

\begin{figure}
  \begin{center}
         \includegraphics[width=0.45\textwidth]{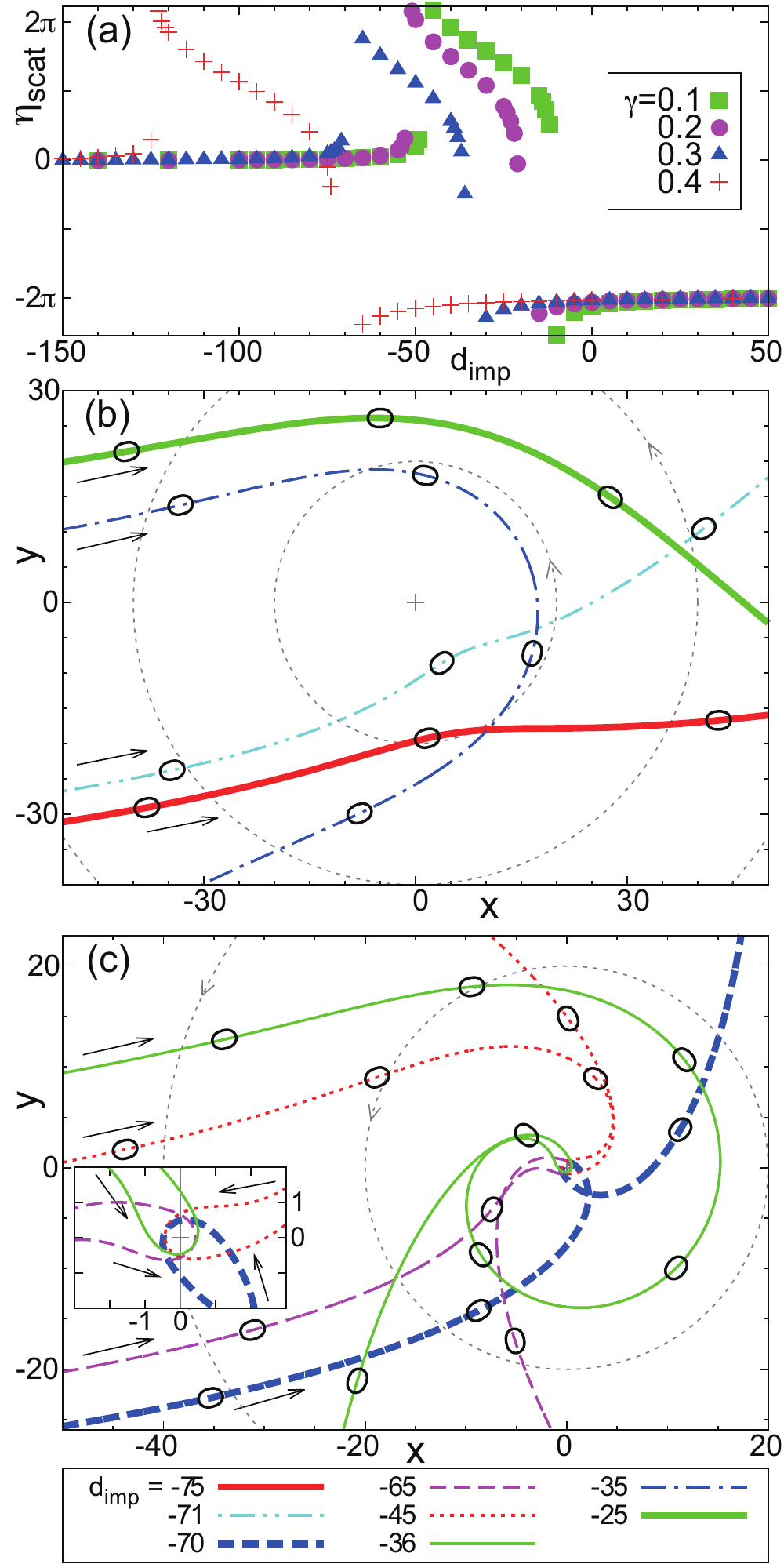}
      \caption{
      Scattering dynamics for $a_1=b_1=+1$ (parallel case): (a) scattering angle $\eta_{\rm scat}$ as a function of the impact parameter $d_{\rm imp}$ for various propulsion strengths $\gamma$; 
      (b,c) example trajectories for $\gamma=0.3$ and for different impact parameters $d_{\rm imp}$.
      Black superimposed silhouettes indicate the swimmer orientations and degrees of deformation. 
      Arrows mark the direction of motion. 
      In addition, the gray dotted lines illustrate the direction of the fluid flow. No capturing events are observed in this case, but the orbit can come very close to the vortex center. 
      (c) In a finite interval of intermediate impact parameters, the swimmer gets transiently caught and loops around the vortex center. As highlighted by the inset, these looping trajectories in close vicinity to the vortex center always show an identical sense of curvature prescribed by the rotational sense of the swirl flow.
      (Other parameter values are $\kappa=0.5$, $\nu_1=1$, and $\mu=1$.)
      } \label{fig:parallel}
  \end{center}
\end{figure}

While they are heading towards the vortex, 
the situation for active swimmers of parallel configuration is just the other way around as for those of perpendicular configuration; 
their trajectory gets curved into the opposite direction during this initial process, see Fig.~\ref{fig:incident_trajectory}. 
Therefore significant scattering now already takes place for negative impact parameters $d_{\rm imp}$ as demonstrated in Fig.~\ref{fig:parallel}. 

First, for very negative impact parameters $d_{\rm imp}$, the swimming trajectory is only slightly influenced by the swirl. The swimmer passes the vortex with only little net change in the propulsion direction, i.e., $\eta_{\rm scat}$ is relatively small, as displayed in Figs.~\ref{fig:incident_trajectory}(b), \ref{fig:parallel}(a), and \ref{fig:parallel}(b) for $\gamma=0.3$ and $d_{\rm imp}=-75$. 
Again, the scattering angle appears higher in Fig.~\ref{fig:incident_trajectory}(b) due to the rescaled $y$ dimension; see Fig.~\ref{fig:parallel}(a) for the absolute value.
Increasing the impact parameter, the swimmer comes closer to the vortex center and the scattering angle $\eta_{\rm scat}$ increases, see the trajectory for $d_{\rm imp}=-71$ in Fig.~\ref{fig:parallel}(b).  

Interestingly, we observe a discontinuous jump of the scattering angle to values $\eta_{\rm scat} > 2 \pi$ in Fig.~\ref{fig:parallel}(a) at higher impact parameters. 
The trajectory for $d_{\rm imp}=-70$ in Fig.~\ref{fig:parallel}(c) shows the drastic event that occurs in this case and explains the jump in the scattering angle. The swimmer gets transiently caught by the swirl. Its trajectory describes a loop of more than a full rotation around the vortex center, before the swimmer can escape with a net scattering angle $\eta_{\rm scat} > 2 \pi$. As indicated by the trajectory for $d_{\rm imp}=-65$ in Fig.~\ref{fig:parallel}(c), this behavior persists for further increasing impact parameters. However, the net scattering angle in Fig.~\ref{fig:parallel}(a) decreases and the loop around the vortex center does not describe a complete rotation of $2\pi$ any more. 

Remarkably, despite the continuous decrease in the net scattering angles in Fig.~\ref{fig:parallel}(a), a qualitative difference appears in the trajectories for higher impact parameters. As illustrated for $d_{\rm imp}=-45$ and $-36$ in Fig.~\ref{fig:parallel}(c), the swimmer now first passes the vortex center on the opposite side. Still, however, the swimmer is transiently caught by the swirl and describes a loop around the vortex center as shown in Fig.~\ref{fig:parallel}(c). As highlighted by the inset of Fig.~\ref{fig:parallel}(c), the swimmer in all these cases always performs the loop around the center with the same rotational sense as the fluid flow. This is true independently of the side on which the swimmer first passes the vortex center. 
Consequently, in the latter two cases of $d_{\rm imp}=-45$ and $-36$, the swimmer must switch the side that it exposes to the center of the swirl. 

Finally, there is another discontinuous jump of the scattering angle in Fig.~\ref{fig:parallel}(a) at still higher impact parameters. Corresponding trajectories in Fig.~\ref{fig:parallel}(b) for $d_{\rm imp}=-35$ and $-25$ reveal the reason for this jump. The swimmer does not perform a narrow loop around the vortex center any more. Instead its trajectory features a simple bend around the swirl. We observe events from close to backscattering up to practically no net scattering at all for large impact parameters. Again, for clarity, we shift the scattering angles corresponding to these events by $-2\pi$ in Fig.~\ref{fig:parallel}(a). 
The complete trajectory of the scattering event for $d_{\rm imp}=-25$ is depicted in Fig.~\ref{fig:incident_trajectory}(b).

\subsection{Effect of thermal noise} \label{sec:noise}

Finally, we ask the question whether thermal noise \cite{Romanchuk_review,Howse2007,van_Teeffelen_Loewen_PRE_2008,BtT2011} can qualitatively modify the above results on the scattering and capturing dynamics. In particular, we test the stability of the capturing event against noise and analyze whether the trajectories of the captured state are stable against thermal fluctuations. 
For this purpose, we add a stochastic force term $\bm{\xi}$ to the dynamic equation of relative velocity, Eq.~(\ref{eq:v}). 
To keep our argument simple, we consider Gaussian white noise characterized by $\langle{\xi}_{i}(t)\rangle = {0}$ and $\langle \xi_{i}(t) \xi_{j}(0)\rangle = \sigma^2 \delta_{ij} \delta(t)$. 
Here $\sigma$ quantifies the strength of the stochastic fluctuations. 

Under these conditions we now repeat the scattering experiment that led to the captured states for the perpendicular configuration. The parameters are set to the same values as before (see the caption of Fig.~\ref{fig:perpendicular}, $\gamma=0.3$). We start from the same large initial distance $r_{\rm init}=1.5 \times 10^4$ from the vortex center. However, besides the impact parameter $d_{\rm imp}$, we now vary the noise intensity $\sigma$. 

\begin{figure}
  \begin{center}
         \includegraphics[width=0.4\textwidth]{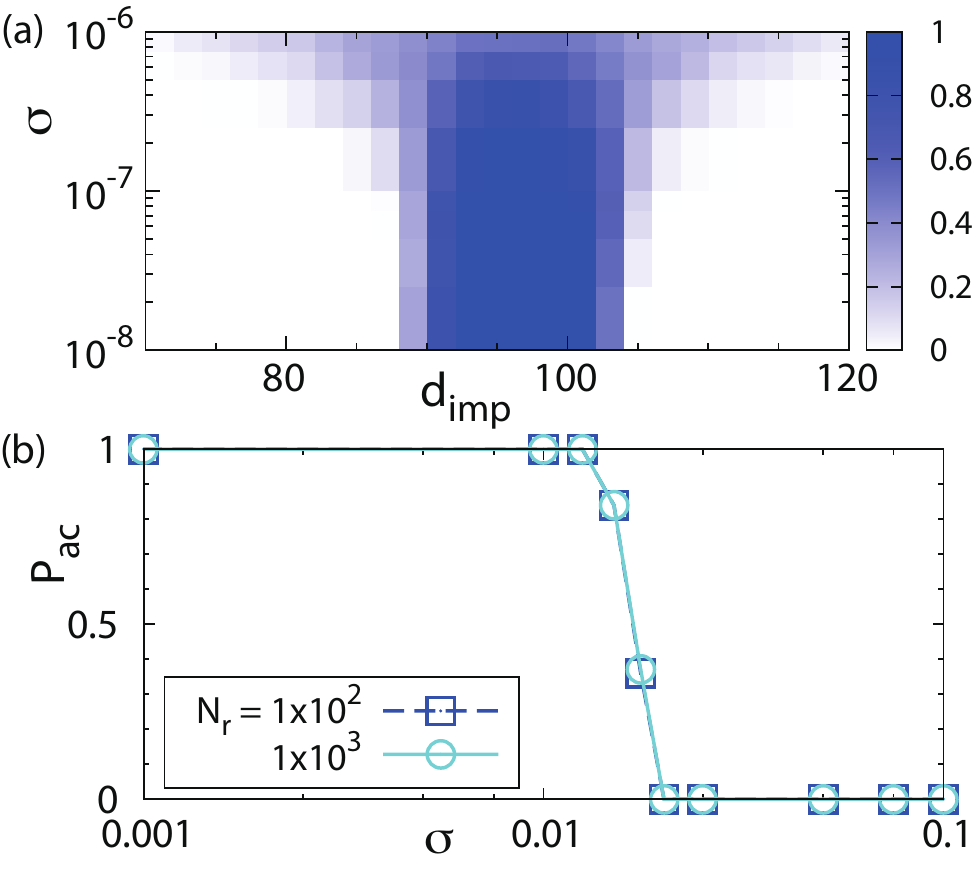}
      \caption{
      Effect of thermal noise on the scattering dynamics. We test the stability of the capturing event for a swimmer in the perpendicular configuration that is characterized by the same parameter values as in Fig.~\ref{fig:perpendicular} ($\gamma=0.3$). 
      (a) Probability for the swimmer to get captured by the swirl when heading from an initial distance $r_{\rm init} = 1.5\times 10^4$ towards the vortex center. The probability is plotted as a function of the impact parameter $d_{\rm imp}$ and the strength $\sigma$ of the thermal noise. For the color code see the scale bar on the right. 
      (b) Probability $P_{\rm ac}$ of a captured swimmer 
      to still remain on a captured active circular trajectory after $N_r$ circulations, plotted as a function of the stochastic noise strength $\sigma$. 
      } \label{fig:noise_perpendicular}
  \end{center}
\end{figure}

Due to the influence of the stochastic force, we now have to measure the probability for the swimmer to get captured. 
For each impact parameter $d_{\rm imp}$ and noise intensity $\sigma$, we thus counted the number of capturing and scattering events for repeated runs with different realizations of the stochastic noise. The probabilities were determined from the overall statistics and are shown in Fig.~\ref{fig:noise_perpendicular}(a). 

At weak noise intensities $\sigma\approx10^{-8}$, the stochastic effects are negligible and capturing events occur approximately in the same way as in the deterministic case. To interpret the noise strengths, we use Stokes' relation and Einstein's relation, and we further assume that the thermal noise for active swimmers is of the same magnitude as in the passive case. We then can attribute the thermal noise of strength $\sigma\approx10^{-8}$ to a spherical self-propelled droplet of a few millimeters in size in an aqueous solution. 
At noise intensities $\sigma\approx10^{-7}$, corresponding to a swimmer size around one millimeter, the capturing event is still well defined as a function of the impact parameter $d_{\rm imp}$. 
In contrast to that, at a noise strength $\sigma\approx10^{-6}$, the stochastic fluctuations smear out the statistics and a capturing event is not well-defined any more. 
This intensity of thermal noise identifies a spherical object of a few hundred micrometers in size, which comes into the range of large microorganisms. 

The reason for the fading of the capturing event in the latter case is not related to the captured states becoming unstable, however. This can be seen by considering captured particles on their active circular motion around the vortex center and exposing them to the stochastic force. We measured the probability $P_{\rm ac}$ of finding the particles still captured after a certain number $N_r$ of circulations around the swirl. In practice, we checked whether they are still within a certain threshold distance from the vortex center. Our results are summarized in Fig.~\ref{fig:noise_perpendicular}(b). 

It is obvious that the swimmers can escape from the captured states only at much higher noise intensities than the ones referred to in Fig.~\ref{fig:noise_perpendicular}(a). The fading probabilities of getting captured at higher noise strengths $\sigma$ in Fig.~\ref{fig:noise_perpendicular}(a) therefore do not imply that the captured states become unstable. Rather, due to the noise, the swimmers increasingly depart from their deterministic trajectories 
towards the swirl center and do not hit the vortex any more. This can easily be cured by placing the swimmers closer to the vortex center in a modified initial condition. Therefore, despite thermal noise, the microswimmers should show the predicted dynamics.

\section{Summary and conclusion} \label{sec:discussion}

In summary, we studied the dynamic behavior of a deformable microswimmer in a swirl flow. To our knowledge this geometry has not been investigated before for active microswimmers, although it is a setup of high practical relevance and straightforward to be realized in an experiment. 
Within our framework, we distinguished between two types of swimmers: 
those that tend to elongate perpendicularly to their propulsion direction and those that prefer a parallel configuration. 
Considering droplets propelling due to chemical reactions, a recent theoretical study suggests that the first ones can be classified as pushers while the second ones are pullers \cite{Yoshinaga2013}. 

For the different types of swimmers, we found different bound states in the swirl. 
Passive deformable swimmers show \textit{passive circular motion}, where not all radii are stable depending on the friction with the fluid environment. Active deformable swimmers at lower strength of self-propulsion and in the perpendicular configuration either escape from the swirl, or they feature an \textit{active circular motion} of one stable radius determined by their activity. In the parallel configuration they always escape from the swirl. Beyond a threshold for the strength of self-propulsion, the swimmers could not escape any more but showed an \textit{active circular motion}, a \textit{lunar-type motion}, or a \textit{multi-circular motion}, depending on the swimmer configuration and on the initial conditions.

Second, in analogy to classical scattering experiments, we investigated how active swimmers interact with the swirl when they are heading towards the vortex from far away in a straight motion. 
For active swimmers of perpendicular configuration we observed that they were \textit{captured} by the swirl on a trajectory of active circular motion at intermediate impact parameters. 
This capturing event is \textit{reentrant} as a function of the strength of self-propulsion. For other impact parameters the swimmer is \textit{scattered} and manages to escape. The scattering event is \textit{reentrant} as a function of the impact parameter and \textit{twofold reentrant} as a function of the strength of self-propulsion. 
In contrast to that, we observed that active swimmers in the parallel configuration are \textit{always scattered}. Thus, to design an active deformable swimmer that is not captured by swirls, the parallel configuration should be preferred and the strength of self-propulsion should be weak enough to avoid circular motions. Nevertheless, scattered swimmers may perform interim loops around the vortex center at very close distances, possibly changing the side that they expose to the vortex center. 
Additional thermal noise was not found to qualitatively alter the capturing dynamics in an appropriately chosen setup. 

The swirl flow geometry is of high practical relevance and straightforward to be realized in an experiment. In principle, a magnetic stir bar on the bottom of a water tank is enough to create the vortex flow. Self-driven droplets on the water surface constitute appropriate deformable active swimmers \cite{Nagai2005,Takabatake2011} 
that propel in a quasi-two-dimensional environment. Therefore they should immediately allow us to test and verify our predictions in an actual experimental setup. This system should be easier to realize than the typically investigated flow profiles of linear shear. 
Apart from that, also on the theoretical side several questions directly follow from our study. 
Most notably, it will be worthwhile also to analyze the swimmer behavior in flow fields of non-vanishing local vorticity or even in turbulent flows. 
On the one hand, hydrodynamically enforced trapping and particle segregation can occur for passive particles in a vorticity flow \cite{Lohse2008,Hanggi2013}. On the other hand, microswimmers were observed to create turbulence-like flows even by themselves in dense suspensions \cite{WensinkPNAS,Wensink2012,Dunkel2013}. 
Apart from that, it has been reported that active swimmers not only self-propel but also actively rotate \cite{Tarama2012,Tarama2013PTEP,Tarama2013PRE,Wittkowski2012,tenHagen_PRE_2011,Zoettl_Stark_PRL_2012,Fily-Baskaran-Marchetti2012,Glotzer2013,Uchida2010,Uchida2011,Kummel2013,Kapral2010,Takagi2013, Pagonabarraga2008,Mallouk2009,Howse2010,Kapral2010_2,Takabatake2011}. 
It will be very interesting to see how the active rotations interact with the rotational component of the flow fields and what their impact is on the dynamics. 
We thus hope that our results will stimulate further investigations both on the theoretical and the experimental side to elucidate the dynamics of active microswimmers in external flow fields.

\section*{Acknowledgements}

A.M.M. and H.L. thank the Deutsche Forschungsgemeinschaft for support of this work through the priority program on microswimmers SPP 1726. 
M.T. acknowledges JSPS for a JSPS Research Fellowship. 
This work was also supported by the JSPS Core-to-Core Program ``Non-equilibrium dynamics of soft matter and information''.

\appendix*

\section{Stability analysis} \label{App:stability analysis}

In this Appendix, we describe the details of the stability analysis of the steady-state solutions of Eqs.~(\ref{eq:x})--(\ref{eq:S}).
First, for the vortex flow specified by Eq.~(\ref{eq:u}), the strain rate tensor of the flow field follows via Eq.~(\ref{eq:A}). At the particle position $\mathbf{x}$, parameterized by Eq.~(\ref{eq:x2d}), we obtain
\begin{equation}
\mathbf{A} = 
 \left(
 \begin{array}{cc}
 \mu r^{-2} \sin 2\eta & -\mu  r^{-2} \cos 2\eta \\[.1cm]
 -\mu r^{-2} \cos 2\eta & - \mu  r^{-2} \sin 2\eta
 \end{array}
 \right).
 \label{eq:A2d}
\end{equation}
Inserting it together with the parameterizations Eqs.~(\ref{eq:x2d})--(\ref{eq:S2d}) into Eqs.~(\ref{eq:x})--(\ref{eq:S}), the equations of motion can be rewritten in the form 
\begin{equation}
\frac{d r}{d t} = v \cos (\Theta-\Psi),
 \label{eq:2d.r}
\end{equation}
\begin{equation}
\frac{d v}{d t} = \gamma v -v^3  -a_1 v s \cos 2\Psi,
 \label{eq:2d.v}
\end{equation}
\begin{equation}
\frac{d s}{d t} = -\kappa s +\frac{b_1}{2} v^2 \cos 2\Psi - \nu_1 \mu r^{-2} \sin 2 \Theta,
 \label{eq:2d.s}
\end{equation}
\begin{eqnarray}
\frac{d \Theta}{d t} 
 &=& -\frac{b_1}{4 s} v^2 \sin 2 \Psi - v r^{-1}  \sin (\Theta-\Psi) \notag \\
 &&  \quad{}-\mu r^{-2} \left( 1 +\frac{\nu_1}{2 s} \cos 2\Theta \right), 
 \label{eq:2d.Theta}
\end{eqnarray}
\begin{equation}
\frac{d \Psi}{d t} = \left( a_1 s -\frac{b_1}{4 s} v^2 \right) \sin 2 \Psi - \frac{\nu_1}{2 s} \mu r^{-2} \cos 2 \Theta.
 \label{eq:2d.Psi}
\end{equation}
We here defined $\Theta = \theta-\eta$ and $\Psi = \theta -\phi$. 

Following the general procedure, we investigate the stability of the steady-state solutions of these equations via the eigenvalues of the corresponding linear stability matrix. Its components are defined by
\begin{equation}
\mathcal{L}_{ij} = \frac{\partial}{\partial{\mathcal{X}_j}} \left( \frac{d \mathcal{X}_i}{d t} \right),
 \label{eq:L}
\end{equation}
where $\bm{\mathcal{X}} = \left( r, v, s, \Theta, \Psi \right)$.
We obtain the eigenvalues $\lambda$ of $\bm{\mathcal{L}}$ as usual from the condition
\begin{equation}
\det \left( \bm{\mathcal{L}} -\lambda \bm{\mathcal{I}} \right) = 0,
 \label{eq:2d.21}
\end{equation}
with $\bm{\mathcal{I}}$ the unity matrix. The corresponding motion is stable, if all $\lambda<0$; marginally stable, if all $\lambda\leq0$; and it becomes unstable, if at least one $\lambda>0$. 

From Eq.~(\ref{eq:2d.v}), two types of steady-state solutions follow. One of them describes a passive motion of $v = 0$, i.e.\ the particle is simply advected by the fluid flow. The other one corresponds to an active motion given by the relative speed $v = \sqrt{ \Gamma}$ with respect to the surrounding fluid, where
\begin{equation}
\Gamma = \gamma - a_1 s \cos 2 \Psi.
 \label{eq:Gamma}
\end{equation}
In the remaining part of this Appendix, we carry out the linear stability analysis of the passive motion. 

For the passive motion $v=0$, Eq.~(\ref{eq:2d.r}) implies that the distance $r$ from the vortex center remains constant. We thus obtain circular trajectories of fixed radius $r = r_0$ in the passive case. This is why we term this kind of motion the \textit{passive circular motion}. 
The complete steady-state solution is found by setting the remaining time derivatives in the above dynamic equations equal to zero. 
Taking into account that $v=0$, it then follows from Eq.~(\ref{eq:2d.s}) that 
\begin{equation}
s = -\frac{\nu_1}{\kappa} \mu r_0^{-2} \sin 2\Theta.
 \label{eq:2d.1}
\end{equation}
Likewise, from Eq.~(\ref{eq:2d.Theta}) together with Eq.~(\ref{eq:2d.s}), we obtain 
\begin{equation}
\tan 2\Theta = \frac{\kappa r_0^{2}}{2 \mu }.
 \label{eq:2d.2}
\end{equation} 
The dynamic equation Eq.~(\ref{eq:2d.Psi}) can be ignored at this point because it determines the relative orientation of the relative velocity $\mathbf{v}$, which vanishes in the case of passive circular motion $v=0$. 
Naturally, it becomes important in the following when we study the bifurcation from the passive circular motion ($v=0$) to other types of motion characterized by $v \neq 0$. 

We determined the eigenvalues of the linear stability matrix for the passive circular motion via Eq.~(\ref{eq:2d.21}). They are obtained as
$\lambda_r = 0$, $\lambda_v = \Gamma$, $\lambda_{\Psi} = 2 a_1 s \cos 2 \Psi$, and as the eigenvalues $\lambda_{\pm}$ of the submatrix
\begin{equation}
\bm{\mathcal{L}}^{\rm sub} = 
 \left(
 \begin{array}{cc}
 -\kappa & -2 \nu_1 \mu r_0^{-2} \cos 2\Theta \\[.1cm]
 \frac{1}{2} \nu_1 \mu r_0^{-2} s^{-2} \cos 2\Theta & \nu_1 \mu r_0^{-2}s^{-1} \sin 2\Theta
 \end{array}
 \right).
 \label{eq:Ldag}
\end{equation}
On the one hand, since $s>0$ (being the magnitude of deformation), we conclude from Eq.~(\ref{eq:2d.1}) that $\nu_1\mu\sin 2\Theta <0$.
Consequently, $\tr \bm{\mathcal{L}}^{\rm sub} <0$ and $\det \bm{\mathcal{L}}^{\rm sub} >0$. 
This leads to $\lambda_{\pm}<0$, 
which is necessary for stability. 
On the other hand, the eigenvalue $\lambda_r=0$ implies that the stability is at most marginal. 
However, we still need to consider the signs of the eigenvalues $\lambda_v$ and $\lambda_{\Psi}$. 
Both, $\lambda_v$ and $\lambda_{\Psi}$, depend on $\Psi$, so we now take into account Eq.~(\ref{eq:2d.Psi}). 

In the case of passive circular motion, 
the steady-state solution of Eq.~(\ref{eq:2d.Psi}) follows as 
\begin{equation}
\sin 2\Psi 
 = \frac{\kappa}{a_1 \nu_1 \sin 2\Theta}.
 \label{eq:2d.3}
\end{equation}
This expression does not determine the sign of $\cos 2\Psi$. Thus there is always a solution that guarantees the condition $\lambda_{\Psi} = 2 a_1 s \cos 2 \Psi\leq0$ necessary for marginal stability. 
Nevertheless, we must satisfy $\sin^2 2\Psi \leq 1$ for the steady-state solution to exist. This leads to the condition $r_0 \geq r_{0, {\rm min}}$ with $r_{0, {\rm min}}$ given by 
\begin{equation}
r_{0, {\rm min}} = \left( 2 |\mu| \right)^{1/2} \left\{ \left( a_1 \nu_1 \right)^2 -\kappa^2 \right\}^{-1/4}. 
 \label{eq:r0min}
\end{equation}
Taking into account Eqs.~(\ref{eq:Gamma}) and (\ref{eq:2d.1}) together with the last eigenvalue $\lambda_v=\Gamma$, we need to require 
\begin{equation}
\lambda_v  = \Gamma = \gamma +a_1 \cos 2\Psi  \,\frac{\nu_1 \mu r_0^{-2}}{\kappa}  \sin 2\Theta  < 0
 \label{eq:2d.6}
\end{equation}
for the solution $v=0$ to be stable. 
Together with Eqs.~(\ref{eq:2d.3}) and (\ref{eq:2d.2}), we obtain $\gamma < \gamma_-$ with $\gamma_-$ given by 
\begin{equation}
\gamma_- 
 = -\frac{2 \mu^2}{ r_0^{2} r_{0, {\rm min}}^2} \left\{ \frac{  \left( r_0^4  -r_{0, {\rm min}}^4 \right) }{ \kappa^2  r_0^{4} +4 \mu^2} \right\}^{1/2}<0.
 \label{eq:gamma-}
\end{equation}
When $r_0$ approaches $r_{0, {\rm min}}$ from above, we find that $\gamma_-$ tends to zero. 
 
Finally, when $r_0 < r_{0, {\rm min}}$, the steady-state solution for $\Psi$ does not exist. Eq.~(\ref{eq:2d.Psi}) then implies that $\Psi$ monotonically increases or decreases, depending on the parameters. 
Then, on average, $\cos 2\Psi$ vanishes, 
and Eq.~(\ref{eq:2d.6}) reduces to $\gamma < 0$ for $r_0 < r_{0, {\rm min}}$. 
We tested and confirmed this observation numerically (see also Fig.~\ref{fig:gamma-vsr0}). 
Our results are summarized by the necessary condition 
\begin{equation}
\gamma <
\left\{
\begin{array}{cc}
0 &~~~{\rm for}~ r_0 < r_{0, {\rm min}} \\
\gamma_- &~~~{\rm for}~ r_0 \ge r_{0, {\rm min}}
\end{array}
\right.
 \label{eq:stability_passive_circular}
\end{equation}
for the passive circular motion to be marginally stable. 

Eqs.~(\ref{eq:r0min}), (\ref{eq:gamma-}), and (\ref{eq:stability_passive_circular}) imply that for an increasing stiffness of the particle (i.e.\ increasing $\kappa$) the stability range of the passive circular steady-state solution increases. This can be seen as follows. 
As $\kappa$ approaches $|a_1\nu_1|$ from below, the value of $r_{0, {\rm min}}$ diverges. For $\kappa\geq|a_1\nu_1|$, the condition for $\gamma$ in Eq.~(\ref{eq:stability_passive_circular}) extends to the full range of $\gamma<0$. 
This corresponds to the natural requirement that the passive particle suffers from friction with its fluid environment, with $\gamma<0$ setting the friction parameter. 
Thus, in this case, circular steady-state trajectories of all radii are marginally stable.


\begin{thebibliography}{15}

\bibitem{Cates_review}
M. E. Cates, Rep. Prog. Phys. {\bf 75}, 042601 (2012).

\bibitem{Romanchuk_review}
P. Romanczuk, M. B\"ar, W. Ebeling, B. Lindner, and L. Schimansky-Geier, Eur. Phys. J. Special Topics {\bf 202}, 1 (2012).

\bibitem{Marchetti_review}
M. C. Marchetti, J. F. Joanny, S. Ramaswamy, T. B. Liverpool, J. Prost, M. Rao, and R. A. Simha, Rev. Mod. Phys. {\bf 85}, 1143 (2013).

\bibitem{Berke_Lauga_PRL_2008}
A. P. Berke, L. Turner, H. C. Berg, and E. Lauga, Phys. Rev. Lett. {\bf 101}, 038102 (2008). 

\bibitem{Wensink_Loewen_PRE_2008}
H. H. Wensink and H. L\"owen, Phys. Rev. E {\bf 78}, 031409 (2008).

\bibitem{van_Teeffelen_Loewen_PRE_2008}
S. van Teeffelen and H. L\"owen, Phys. Rev. E {\bf 78}, 020101 (2008).

\bibitem{Zimmermann_Teeffelen_Soft_Matter_2009}
S. van Teeffelen, U. Zimmermann, and H. L{\"o}wen, Soft Matter {\bf 5}, 4510 (2009). 

\bibitem{Takagi_Palacci_archive}
D. Takagi, J. Palacci, A. B. Braunschweig, M. J. Shelley, J. Zhang, 
Soft Matter {\bf 10}, 1784 (2014). 

\bibitem{Gompper2013}
J. Elgeti and G. Gompper, Europhys. Lett. {\bf 101}, 48003 (2013). 

\bibitem{Stocker_PRL}
P. Peruzzo, A. Defina, H. M. Nepf, and R. Stocker, Phys. Rev. Lett. {\bf 111}, 164501 (2013). 

\bibitem{Peyla_2013}
X. Garcia, S. Rafa\"i, and P. Peyla, Phys. Rev. Lett. {\bf 110}, 138106 (2013). 

\bibitem{Zoettl_Stark_PRL_2012}
A. Z\"ottl and H. Stark, Phys. Rev. Lett. {\bf 108}, 218104 (2012).

\bibitem{tenHagen_PRE_2011}
B. ten Hagen, R. Wittkowski, and H. L\"{o}wen, Phys. Rev. E {\bf 84}, 031105 (2011). 

\bibitem{Baraban}
L. Baraban, D. Makarov, O. G. Schmidt, G. Cuniberti, P. Leiderer, and A. Erbe, Nanoscale {\bf 5}, 1332 (2013).

\bibitem{Aranson_PNAS}
A. Sokolov, M. M. Apodaca, B. A. Grzybowski, and I. S. Aranson, Proc. Natl. Acad. Sci. USA {\bf 107}, 969 (2010).

\bibitem{di_Leonardo_PRL}
R. Di Leonardo, E. Cammarota, G. Bolognesi, H. Sch\"afer, and M. Steinhart, Phys. Rev. Lett. {\bf 107}, 044501 (2011). 

\bibitem{di_Leonardo_NJP}
L. Angelani and R. Di Leonardo, New J. Phys. {\bf 12}, 113017 (2010). 

\bibitem{Juelicher}
F. J\"ulicher, A. Ajdari, and J. Prost, Rev. Mod. Phys. {\bf 69}, 1269 (1997).

\bibitem{Tarama2013JCP}
M. Tarama, A. M. Menzel, B. ten Hagen, R. Wittkowski, T. Ohta, and H. L\"owen, J. Chem. Phys. {\bf 139}, 104906 (2013). 

\bibitem{Nagai2005}
K. Nagai, Y. Sumino, H. Kitahata, and K. Yoshikawa, Phys. Rev. E {\bf 71},
065301 (2005).

\bibitem{Takabatake2011}
F. Takabatake, N. Magome, M. Ichikawa, and K. Yoshikawa, J. Chem.
Phys. {\bf 134}, 114704 (2011).

\bibitem{Tarama2012}
M. Tarama and T. Ohta, J. Phys.: Condens. Matter {\bf 24}, 464129 (2012). 

\bibitem{Tarama2013PTEP}
M. Tarama and T. Ohta, Prog. Theor. Exp. Phys., 013A01 (2013). 

\bibitem{Tarama2013PRE}
M. Tarama and T. Ohta, Phys. Rev. E {\bf 87}, 062912 (2013). 

\bibitem{Wittkowski2012}
R. Wittkowski and H. L\"owen, Phys. Rev. E {\bf 85}, 021406 (2012). 

\bibitem{Fily-Baskaran-Marchetti2012}
Y. Fily, A. Baskaran, and M. C. Marchetti, Soft Matter {\bf 8}, 3002 (2012). 

\bibitem{Glotzer2013}
N. H. P. Nguyen, D. Klotsa, M. Engel, and S. C. Glotzer, 
Phys. Rev. Lett. {\bf 112}, 075701 (2014). 

\bibitem{Uchida2010}
N. Uchida and R. Golestanian, Phys. Rev. Lett. {\bf 104}, 178103 (2010). 

\bibitem{Uchida2011}
N. Uchida and R. Golestanian, Phys. Rev. Lett. {\bf 106}, 058104 (2011). 

\bibitem{Kummel2013}
F. K\"ummel, B. ten Hagen, R. Wittkowski, I. Buttinoni, R. Eichhorn, G. Volpe, H. L\"owen and C. Bechinger, Phys. Rev. Lett. {\bf 110}, 198302 (2013). 

\bibitem{Kapral2010}
S. Thakura and R. Kapral, J. Chem. Phys. {\bf 133}, 204505 (2010). 
\bibitem{Takagi2013}
D. Takagi, A. B. Braunschweig, J. Zhang, and M. J. Shelley, Phys. Rev. Lett. {\bf 110}, 038301 (2013). 

\bibitem{TonerTuPRL}
J. Toner and Y. Tu, Phys. Rev. Lett. {\bf 75}, 4326 (1995). 
\bibitem{TonerTuPRE}
J. Toner and Y. Tu, Phys. Rev. E {\bf 58}, 4828 (1998). 
\bibitem{TonerTuRamaswamy}
J. Toner, Y. Tu, and S. Ramaswamy, Ann. Phys. 318, 170 (2005). 

\bibitem{OOS2009}
T. Ohta, T. Ohkuma, and K. Shitara, Phys. Rev. E {\bf 80}, 056203 (2009).

\bibitem{degennes}
P. G. de Gennes and J. Prost, {\it The Physics of Liquid Crystals} (Clarendon Press, Oxford, 1993). 

\bibitem{Maffettone1998}
P. L. Maffettone and M. Minale, J. Non-Newtonian Fluid Mech. {\bf 78}, 227 (1998).

\bibitem{OhtaOhkuma2009}
T. Ohta and T. Ohkuma, Phys. Rev. Lett. {\bf 102}, 154101 (2009).  

\bibitem{Shitara2011}
K. Shitara, T. Hiraiwa, and T. Ohta, Phys. Rev. E {\bf 83}, 066208 (2011).

\bibitem{Hiraiwa2010}
T. Hiraiwa, M. Y. Matsuo, T. Ohkuma, T. Ohta, and M. Sano, Europhys. Lett. {\bf 91}, 20001 (2010). 

\bibitem{Hiraiwa2011}
T. Hiraiwa, K. Shitara, and T. Ohta, Soft Matter {\bf 7}, 3083 (2011). 

\bibitem{Tarama2011}
M. Tarama and T. Ohta, Eur. Phys. J. B {\bf 83}, 391 (2011). 

\bibitem{Itino2011}
Y. Itino, T. Ohkuma, and T. Ohta,  J. Phys. Soc. Jpn. {\bf 80},  033001 (2011).

\bibitem{Itino2012}
Y. Itino and T. Ohta, J. Phys. Soc. Jpn. {\bf 81}, 104007 (2012). 

\bibitem{Menzel2012}
A. M. Menzel and T. Ohta, Europhys. Lett. {\bf 99}, 58001 (2012). 

\bibitem{epjst}
M. Tarama, Y. Itino, A. M. Menzel, and T. Ohta, Eur. Phys. J. Special Topics 
{\bf 223}, 121 (2014).

\bibitem{Yoshinaga2013}
N. Yoshinaga, 
Phys. Rev. E {\bf 89}, 012913 (2014). 

\bibitem{Howse2007}
J. R. Howse, R. A. L. Jones, A. J. Ryan, T. Gough, R. Vafabakhsh, and R. Golestanian, Phys. Rev. Lett. {\bf 99}, 048102 (2007).

\bibitem{BtT2011}
B. ten Hagen, S. van Teeffelen, and H. L\"owen, J. Phys.: Condens. Matter {\bf 23}, 194119 (2011).

\bibitem{Lohse2008}
E. Calzavarini, M. Cencini, D. Lohse, and F. Toschi, Phys. Rev. Lett. {\bf 101}, 084504 (2008).

\bibitem{Hanggi2013}
S. Martens, A. V. Straube, G. Schmid, L. Schimansky-Geier, and P. H\"{a}nggi, Phys. Rev. Lett. {\bf 110}, 010601 (2013).

\bibitem{WensinkPNAS}
H. H. Wensink, J. Dunkel, S. Heidenreich, K. Drescher, R. E. Goldstein, H. L\"{o}wen, and J. M. Yeomans, Proc. Natl. Acad. Sci. USA {\bf 109}, 14308 (2012).

\bibitem{Wensink2012}
H. H. Wensink and H. L\"{o}wen, J. Phys.: Condens. Matter {\bf 24}, 464130 (2012).

\bibitem{Dunkel2013}
J. Dunkel, S. Heidenreich, K. Drescher, H. H. Wensink, M. B\"{a}r, and R. E. Goldstein, Phys. Rev. Lett. {\bf 110}, 228102 (2013).

\bibitem{Pagonabarraga2008}
P. Tierno, R. Golestanian, I. Pagonabarraga, and F. Sagu\'{e}s, Phys. Rev. Lett. {\bf 101}, 218304 (2008).

\bibitem{Mallouk2009}
Y. Wang, S.-t. Fei, Y.-M. Byun, P. E. Lammert, V. H. Crespi, A. Sen, and T. E. Mallouk, J. Am. Chem. Soc. {\bf 131}, 9926 (2009). 

\bibitem{Howse2010}
S. Ebbens, R. A. L. Jones, A. J. Ryan, R. Golestanian, and J. R. Howse, Phys. Rev. E {\bf 82}, 015304 (2010). 

\bibitem{Kapral2010_2}
L. F. Valadares, Y.-G. Tao, N. S. Zacharia, V. Kitaev, F. Galembeck, R. Kapral, and G. A. Ozin, Small {\bf 6}, 565 (2010).

\end{thebibliography}
\end{document}